\documentclass[showpacs,preprintnumbers,amsmath,amssymb,aps,twocolumn,superscriptaddress]{revtex4}

\def\O{{\cal O} }

\def\A{{\cal A} }

\def\N{{\cal N} }
\def\S{{\cal S} }
\def\L{{\cal L} }

\def\F{{\cal F} }
\def\G {{\cal G} }

\def\dcut{{%
    \setbox0\hbox{D}
    \rlap{\hbox to \wd0{\hss ~/ \hss}}\box0}}

\usepackage{graphicx}
\usepackage{dcolumn}
\usepackage{bm}
\usepackage{dsfont}
\usepackage{amssymb,amsmath}
\usepackage{epsfig}
\usepackage{epstopdf}
\usepackage{latexsym}
\usepackage{graphicx}
\usepackage{subfigure}
\usepackage{booktabs}
\usepackage{color}

\begin{document}

\title{Steady-state physics, effective temperature dynamics in holography}

\author{Arnab Kundu}
\email{arnab.kundu@saha.ac.in}
\affiliation{Theory Group, Department of Physics, University of Texas, Austin, TX 78712}
\affiliation{Theory Division, Saha Institute of Nuclear Physics, 1/AF Bidhannagar, Kolkata 700064, India}

\author{Sandipan Kundu}
\affiliation{Theory Group, Department of Physics, University of Texas, Austin, TX 78712} 
\affiliation{Texas Cosmology Center, University of Texas, Austin, TX 78712}
\affiliation{Department of Physics, Cornell University, Ithaca, New York, 14853, USA}

\preprint{UTTG-19-13, TCC-015-13}

\begin{abstract}
Using the gauge-gravity duality, we argue that for a certain class of out-of-equilibrium steady-state systems in contact with a thermal background at a given temperature, the macroscopic physics can be captured by an effective thermodynamic description. The steady-state is obtained by applying a constant electric field that results in a stationary current flow. Within holography, we consider generic probe systems where an open string equivalence principle and an open string metric govern the effective thermodynamics. This description comes equipped with an effective temperature, which is larger than the background temperature, and a corresponding effective entropy. For conformal or scale-invariant theories, certain scaling behaviours follow immediately. In general, in the large electric field limit, this effective temperature is also observed to obey generic relations with various physical parameters in the system.  
\end{abstract}

\maketitle

\section{Introduction}

Thermodynamics is an integral cornerstone of our understanding of the physical world, where the core principles are based on the existence of a thermal equilibrium. For systems driven out-of-equilibrium, the governing principles are much less understood; see {\it e.g.}~\cite{PhysRevLett.78.2690, PhysRevE.60.2721}. It is extremely difficult to address such questions in a strongly coupled quantum system. First, given a microscopic description, conventional perturbative techniques are inadequate at strong coupling. Second, conceptual insights that may lead to an effective description are also lacking.

In recent years, the AdS/CFT correspondence\cite{Maldacena:1997re}, more generally the gauge-gravity duality\cite{Susskind:1994vu}, has emerged to be an extremely powerful tool to address aspects of strongly coupled physics. String theory provides a large class of concrete examples where this correspondence is precise. Many of these examples correspond to a strongly coupled conformal or a scale-invariant field theory and is naturally equipped in describing quantum criticality. However, criticality is not necessary. The duality works for generic large-$N$ gauge theories with a running coupling constant\cite{Itzhaki:1998dd}, or a confining gauge theory\cite{Klebanov:2000hb}.

In this article, we will use the gauge-gravity duality to explore an emerging principle for a system driven out-of-equilibrium. We will consider a non-equilibrium steady-state (NESS) situation within a probe sector which is kept in contact with a large background at some given temperature $T$. The NESS in the probe sector is induced by introducing a constant external electric field $E$ that drives a constant current. We will argue that {\it all modes} in this probe sector experience an effective temperature, denoted by $T_{\rm eff}$, with respect to which it has a purely thermal behaviour. We will also argue that for conformal systems this effective temperature is always larger than the background temperature: $T_{\rm eff}> T$. By virtue of the probe limit, the heat flow between the probe sector and the background is suppressed. For critical systems, $T_{\rm eff}$ depends only on three ingredients: (i) dimensionality $d$, (ii) the global symmetry group $G$ and (iii) how NESS is induced, {\it i.e.},~$E$. For gapped systems, or systems with a running coupling constant, there may be additional dependences on the beta function of the gauge coupling or other dimensionful parameters in the system.

The existence of the unique effective temperature $T_{\rm eff}$ is, nonetheless, ubiquitous. In generic situations one can verify that $T_{\rm eff} > T$. However, there may be special cases when this is not true, see {\it e.g.}~\cite{Nakamura:2013yqa}. The consequences of this are rather profound: it allows us to define thermodynamics with an effective entropy, obtain a fluctuation-dissipation relation and recover an {\it otherwise thermal} physics for the NESS. In this article, we will discuss the main results for a broader perspective and a technically detailed account will appear elsewhere\cite{Kundu:tobe}.

\section{Holography and Thermal Physics}

In the most familiar example of the AdS/CFT correspondence $G \equiv SO(d,2)$ for a boundary theory in $d$ spacetime dimensions\cite{comment1}. In all known examples, the gravity description emerges from the closed string sector. This is typically described by a $10$-dimensional geometry of the Freund-Rubin type: {\it e.g.}~AdS$_{d+1} \otimes_w M^{9-d}$, where AdS$_{d+1}$ denotes the anti-de Sitter space in $(d+1)$-dimensions and $M^{9-d}$ denotes a compact manifold. The symbol $\otimes_w$ denotes a warped product geometry. This geometry and the various matter fields are sourced by a collection of a large number of branes, which equivalently gives rise to the adjoint sector of the dual large-$N$, $SU(N)$ gauge theory. A finite non-zero temperature is realized as the presence of an event horizon in the geometry. The corresponding Hawking temperature can be obtained by demanding regularity of the Euclidean metric with a compact time direction.

We introduce probe degrees of freedom in the background of a thermal adjoint sector, which transform under the fundamental representation of the $SU(N)$ gauge group. In the full $10$-dimensional geometry, this can be achieved by introducing probe branes of appropriate dimensions, which amounts to introducing open string degrees of freedom in the closed-string background. The details of the probe embedding depend primarily on three factors: (i) dimensionality, (ii) stability of the embedding and (iii) the physics we want to realize. This was pioneered in \cite{Karch:2002sh}. Clearly, the full details of the $10$-dimensional background play a crucial role in such embeddings and each case needs to be studied individually. For simplicity, however, we will abstract away from this detailed constructions and focus on a {\it reduced} description for the probes\cite{note1}.

\subsection{The Probe Embedding and Fluctuations}

Let us begin with a background $(d+1)$-dimensional metric, in the string frame, of the general form
\begin{eqnarray}
ds^2 = g_{tt} dt^2 + g_{xx} dx^2 + g_{uu} du^2 & + & \sum_{i=2}^{d-2} g_{ii} dx^i dx^i \nonumber \\
 & + & ds_{\rm compact}^2    \ , \label{metgen}
\end{eqnarray}
where we assume that the metric functions depend on only one co-ordinate $u$, which is the radial direction. In our convention, $u \to 0$ is the boundary of the bulk spacetime. The dual field theory lives in the $\{t, x, x^i\}$-plane. Furthermore, we collectively represent the compact manifold by $ds_{\rm compact}^2$. In this background we introduce $N_f$ number of probe D$p$-branes which wrap $\{t, x, u\}$ and $m$ of the remaining space-directions (along with $(p-m-2)$ of the compact directions). The dynamics of these probes is determined by the Dirac-Born-Infeld (DBI) action\cite{note2}
\begin{eqnarray}
S_{\rm DBI} & = & - N_f T_p \int d^{p+1} \xi e^{-\Phi} \left( - {\rm det} \left[P[g] +  F \right]\right)^{1/2} , \label{dbi} \\
                    & = & - N_f T_p \int d^{p+1} \xi \, L_{\rm DBI} \ , \\
{\rm with} \quad 1 & = & \left(2 \pi \alpha'\right) \ ,
\end{eqnarray}
where $T_p$ denotes the tension of the D$p$-brane, $\xi$ denotes the collective coordinates on the brane worldvolume (including the compact directions), $\Phi$ is the dilaton field, $P[g]$ denotes the pull-back of the background metric:
\begin{eqnarray}
P[g]_{ab} = \frac{\partial X^\mu}{\partial \xi^a} \frac{\partial X^\nu}{\partial \xi^b} \, g_{\mu\nu} 
\end{eqnarray}
 and $F$ represents a gauge field on the probe worldvolume. For simplicity, we will assume that $P[g]$ is trivial, {\it i.e.}~$\left. P[g] = g \right |_{\rm worldvolume}$, which means that we assume
 \begin{eqnarray}
  X_{(0)}^a = \xi^a  \label{embed}
 \end{eqnarray}
minimizes the probe worldvolume. Here, the subscript $``(0)"$ denotes the classical profile. Note that we have implicitly assumed that the background has vanishing NS-NS two-form. This is chosen for convenience, and a non-vanishing NS-NS two-form will not modify the qualitative physics which we will discuss in this article.

The steady-state physics can be induced by having a gauge field on the worldvolume
\begin{eqnarray}
 A_x & = & - E \, t + a_x(u) , \nonumber \\
 \implies \quad F & = & F^{(0)} = - E \, dt \wedge dx + a_x' \, du \wedge dx \ , \label{classgauge}
 \label{gaugeelectric}
\end{eqnarray}
where $E$ denotes the electric field along $x$-direction and $a_x(u)$ is a function that needs to be determined from the equation of motion. Since the DBI Lagrangian depends only on $F$, it is a functional of $a_x'(u)$  only. Thus we immediately get an integral of motion, which we call $j$
\begin{eqnarray}
\frac{\partial L_{\rm DBI}}{\partial a_x'} = j \ . \label{defj}
\end{eqnarray}
The above definition of $j$ yields the equation of motion for $a_x'$, which can be solved to obtain
\begin{eqnarray}
a_x'^2 = - j^2 \frac{g_{uu} \left( g_{tt} g_{xx} + E^2 \right)} {j^2 g_{tt} + e^{- 2\Phi} g_{tt}^2 \left( \prod_{i=2}^{m+1} g_{ii} \right)} . \label{solh}
\end{eqnarray}
Evidently, the solution for $a_x'$ has a sign ambiguity. This can be fixed by demanding an ``ingoing" boundary condition at the horizon\cite{Alam:2012fw}, which means that the energy-momentum flows into the horizon.
Using this, the on-shell DBI Lagrangian is obtained to be
\begin{eqnarray}
L_{\rm os} = - \frac{\sqrt{g_{tt}}}{e^{2\Phi}} \left( \prod_{i=2}^{m+1} g_{ii} \right) \left[ \frac{-g_{uu} \left(g_{tt} g_{xx} + E^2 \right)}{j^2 + e^{- 2\Phi} g_{tt} \left( \prod_{i=2}^{m+1} g_{ii} \right)} \right]^{1/2} .\label{lagon}
\end{eqnarray}
In general, because of the square-root expressions, the on-shell DBI Lagrangian may not be real-valued for all values of $u$. The reality condition demands that the numerator and the denominator change sign at the same radial location $u_*$. These conditions in turn determine the constant $j$ as a function of $u_*$:
\begin{eqnarray}
&& E = \left.\sqrt{- g_{tt} g_{xx}}\right|_{u=u_*} , \label{ustar} \\
&& j = \left. \sqrt{- g_{tt}} \left( \prod_{i=2}^{m+1} g_{ii} \right)^{1/2} e^{-\Phi} \right|_{u=u_*} . \label{current}
\end{eqnarray}
Here $u_*$ is obtained by solving (\ref{ustar}) and, as observed by the bulk geometry, is a completely unremarkable position. We will see that this $u_*$ will play the key role for the open string degrees of freedom. Following \cite{Karch:2007pd} it can be shown that the constant $j$ is proportional to the boundary current driven by $E$ and consequently we can define a conductivity, $\sigma \equiv j / E$\cite{comment3}.

So far we have discussed the classical profile of the probe brane. Let us now discuss the physics of the fluctuation modes on the probe. From the perspective of the dual field theory, these fluctuations are of three kinds: (i) scalar, (ii) vector and (iii) spinor. The scalar fluctuations are the transverse fluctuations of the probe brane embedding:
\begin{eqnarray}
X^\mu = X_{(0)}^\mu + \varphi^i \, \delta_i^\mu \ ,
\end{eqnarray}
where $X_{(0)}^\mu$ denotes the classical profile of the probe and $\varphi^i$ represents the collective transverse fluctuations. The vector fluctuations correspond to the fluctuations of the classical gauge field on the probe
\begin{eqnarray}
F_{ab} = F^{(0)}_{ab} + \F_{ab} \ ,
\end{eqnarray}
where $F^{(0)}$ is the classical gauge field on the probe in equation (\ref{classgauge}). Finally, the spinors come from a supersymmetric counter-part of the DBI action in (\ref{dbi}), which schematically consists of a standard Volkov-Akulov type term
\begin{eqnarray}
S_{\rm VA} =  - N_f T_p \int d^{p+1} \xi e^{-\Phi} \left( - {\rm det} \left[M + i \bar{\psi} \gamma \nabla \psi \right]\right)^{1/2} , \label{dbi2}
\end{eqnarray}
where $M = P[g] + F$ and the $\gamma$ matrices satisfy anti-commutation relation with respect to $P[g]$:
\begin{eqnarray}
\left\{\gamma_a, \gamma_b\right\} = 2 P[g]_{ab} \ .
\end{eqnarray}
More precisely, the quadratic fluctuation term on a given D$p$-probe brane can be read of from \cite{Marolf:2003ye, Marolf:2003vf, Martucci:2005rb}.

The computation of fluctuation modes around a particular classical brane configuration will involve inverting the matrix $M$, which can then be decomposed into a symmetric and an anti-symmetric part:
\begin{eqnarray}
M^{ab} \equiv \left( \left( P[g] + F^{(0)} \right)^{-1} \right)^{ab}= \S^{ab} + \A^{ab} \ .
\end{eqnarray}
It can be showed that\cite{Seiberg:1999vs}
\begin{eqnarray}
\S^{ab} & = & \left(\frac{1}{P[g]+ F^{(0)}} \cdot P[g] \cdot \frac{1}{P[g]-F^{(0)}}\right)^{ab} \ , \label{osmmet} \\
\S_{ab} & = & P[g]_{ab} - \left( F^{(0)} \cdot P[g]^{-1} \cdot F^{(0)} \right)_{ab} \ , \\
\A^{ab} & = & - \left(\frac{1}{P[g]+ F^{(0)}} \cdot F^{(0)} \cdot \frac{1}{P[g]-F^{(0)}}\right)^{ab} \ .
\end{eqnarray}
The indices are now raised and lowered with respect to the effective metric denoted by $\S$, which also determines the kinetic terms for the corresponding fluctuation modes. Thus, the fluctuations are governed by Lagrangians which take the following schematic form:
\begin{eqnarray}
 \L_{\rm scalar} & = & - \frac{\kappa}{2} e^{- \Phi} \sqrt{-{\rm det} M} \,  \S^{ab} \, \partial_a \varphi^i \,  \partial_b \varphi^i  + \ldots  , \label{fscalar} \\
  \L_{\rm vector} & = & - \frac{\kappa}{4} e^{- \Phi} \sqrt{-{\rm det} M} \, \S^{ab}\S^{cd} \, \F_{ac} \F_{bd} + \ldots , \label{fvector} \\
 \L_{\rm spinor} & = & i \kappa e^{- \Phi} \sqrt{-{\rm det} M} \, \bar{\psi} \, \S^{ab} \tilde{\gamma}_a \nabla_b \psi + \ldots  \ .\label{fspinor}
\end{eqnarray}
In $\L_{\rm spinor}$ we have redefined the gamma matrices by:
\begin{eqnarray}
\tilde{\gamma}^a & = & \left( \S^{ab} + \A^{ab} \right) \gamma_b \nonumber\\
\implies \left\{ \tilde{\gamma}^a , \tilde{\gamma}^b\right\}  & =  & 2 \S^{ab} \ .
\end{eqnarray}
Also in (\ref{fscalar})-(\ref{fspinor}), $\kappa$ denotes an overall constant, the details of which is not relevant for us. The fields $\varphi^i$, $\F$ and $\psi$ represent the various fluctuation modes and the indices $a,b$ represent the worldvolume coordinates on the probe. We have shown only the kinetic parts of the fluctuation Lagrangian; since other potential terms will not affect our conclusions.

The key observation is that all these modes perceive an effective metric, denoted by $\S$, which is different from both the background metric $g$ and the induced metric $P[g]$. This is the so called {\it open string metric}\cite{Seiberg:1999vs}, which governs the dynamics of open string degrees of freedom propagating in a background geometry with an anti-symmetric $2$-form. The open string data, in terms of the closed string data, is given by\cite{Seiberg:1999vs}
\begin{eqnarray}
&& \S_{ab} = P[g]_{ab} - \left(F^{(0)} \cdot P[g]^{-1} \cdot F^{(0)} \right)_{ab} , \label{osm} \\
&& \G_s = g_s \left[\frac{- {\rm det} \left(P[g] + F^{(0)}\right)}{- {\rm det} P[g] }\right]^{1/2} \label{osg}, 
\end{eqnarray}
where $\S$ and $\G_s$ denote the open string metric and the open string coupling; $g$ and $g_s$ denote the closed string metric and the closed string coupling. The structure of the fluctuations suggests an {\it open string equivalence principle}\cite{Gibbons:2001gy}.

For the background in (\ref{metgen}), the embedding in (\ref{embed}) and the gauge field in (\ref{gaugeelectric}), the open string metric is given by
\begin{eqnarray}
ds_{\rm osm}^2 = \S_{tt} \, dt^2 & + & \S_{xx} \, dx^2 + \S_{uu} \, du^2 + 2 \S_{ut} \, du dt  \nonumber\\
& + & \sum_{i=2}^{m+1} \S_{ii} \, dx^i dx^i + {\rm compact} \ ,
\end{eqnarray}
where we are suppressing the compact part on purpose. The various components can be given in terms of the background metric and the gauge field as
\begin{eqnarray}
\S_{tt} & = & g_{tt} + \frac{E^2}{g_{xx}} \ , \quad \S_{uu} = g_{uu} + \frac{a_x'^2}{g_{xx}} \ , \label{Stt} \\
\S_{ut} & = &  - \frac{E}{g_{xx}} a_x' \ , \quad \S_{ii} = g_{ii} \ , \\
\S_{xx} & = & g_{xx} + \frac{a_x'^2}{g_{uu}} + \frac{E^2}{g_{tt}} \ .
\end{eqnarray}
Clearly, the open string metric is non-diagonal in the $\{t, u\}$-plane. We can diagonalize this block by introducing a new coordinate system $\{\tau, u\}$, which is defined as
\begin{eqnarray}
\tau = t + f(u) \ , \quad  f'(u) = \S_{ut}/\S_{tt} \ .
\end{eqnarray}
With this, the open string metric takes the following form
\begin{eqnarray}
ds_{\rm osm}^2 = \S_{tt} \, d\tau^2 + \left(\frac{\S_{uu}\S_{tt} - \S_{ut}^2}{\S_{tt}}\right) du^2 + \ldots , \label{osmred}
\end{eqnarray}
The zeroes of $\S_{tt}$ will now determine the location of an event-horizon of the open string metric. From the definition of $\S_{tt}$ in (\ref{Stt}), we observe that the electric field results in an event horizon of the open string metric at the location $u=u_*$, which is obtained by solving (\ref{ustar}). Subsequently, the open string equivalence principle implies an universal effective temperature in the probe sector. We can read off this temperature by requiring regularity of the Euclidean open string metric and compactifying the $\tau$-direction. Note that, we cannot make Euclidean $g$ and Euclidean $\S$ regular simultaneously since their respective horizons do not coincide, which implies that the background and the probe sector are not in a thermal equilibrium.

Carrying out the usual procedure of Euclideanization of the OSM, periodically identifying the imaginary time-direction and then demanding regularity yields the following formula for the effective temperature
\begin{eqnarray}
T_{\rm eff}  & = & \frac{1}{4\pi} \left ( \sqrt{\frac{E^2 g_{xx}' - g_{xx}^2 g_{tt}'}{- g_{tt} g_{uu} g_{xx}^2}} \right) \nonumber\\
 & \times & \left. \left[ \frac{\left( \prod_{i=2}^{m+1} g_{ii}\right)'}{\left( \prod_{i=2}^{m+1} g_{ii}\right)} g_{tt} + g_{tt}' - 2 \Phi' g_{tt} \right] \right|_{u=u_*} \ ,
\end{eqnarray}
where $(\ldots)' \equiv d / du (\ldots)$. Assuming an $SO(d-1)$ symmetry along the $\mathbb{R}^{d-1}$ directions in the boundary theory, we can choose $g_{ii} = g_{xx} = 1/u^2$. In that case, this effective temperature, given in terms of the background data $\{T, E, \beta\}$, can be recast in the following general form
\begin{eqnarray}
T_{\rm eff} = \left. \frac{1}{4\pi} \left[\frac{\left( 2 E^2 u + g_{tt}'\right) \left(2 E^2 u (m  +  \beta) + g_{tt}'\right)} {- g_{tt} g_{uu}}\right]^{1/2} \right|_{u=u_*}  \label{teff}
\end{eqnarray}
where $\beta(u) := (d \Phi)/ (d\log u)$ represents the beta function corresponding to the running dilaton. Assuming $\beta =0$, from (\ref{teff}) it is clear that the sufficient condition to violate $T_{\rm eff} >T$ is to have $E^2<0$, which will violate weak energy condition for the matter field on the classical probe profile. This is expected intuitively, since maintaining the steady-state continually pumps energy into the system and the resulting $T_{\rm eff}$ should be greater than $T$. In the presence of a non-trivial $\beta$, this inequality $T_{\rm eff} > T$ is not always true\cite{Nakamura:2013yqa}.

The propagation of fluctuations in a DBI-background can be analyzed using the method of Boillat (see {\it e.g.}~\cite{Boillat:1970} and \cite{Gibbons:2000xe} for a more recent discussion). Subsequently, it can be shown that the open string causal structure is determined by the Boillat metric, which belongs to the same conformal class as the open string metric.  Typically, the Boillat light-cone lies inside the Einstein light-cone as long as an weak energy condition is satisfied by the matter field on the probe\cite{Gibbons:2000xe}. This light cone structure is another manifestation of the $T_{\rm eff} >T$ inequality.

This effective temperature is accompanied by an effective entropy as well. In \cite{Alam:2012fw} the Helmholtz free energy for such an NESS probe system was conjectured for a particular system. Before we generalize the proposal in \cite{Alam:2012fw}, let us offer some comments regarding the validity of our analysis. The basic statement of gauge-gravity duality can be succinctly phrased as below:
\begin{eqnarray}
{\cal Z}_{\rm bulk} = {\cal Z}_{\rm QFT} \ ,
\end{eqnarray}
where ${\cal Z}_{\rm bulk}$ corresponds to the {\it bulk gravity path integral} and ${\cal Z}_{\rm QFT}$ is the field theory path integral. The quantity ${\cal Z}_{\rm bulk}$ is defined as
\begin{eqnarray}
{\cal Z}_{\rm bulk} & = & e^{i S_{\rm bulk}} \ , \\
S_{\rm bulk} & = & S_{\rm grav} + S_{\rm probe} \ .
\end{eqnarray}
Here $S_{\rm grav}$ is the $10$ or $11$-dimensional supergravity action which gives rise to the metric solution in (\ref{metgen}). Generically, form fields of various ranks will also be present in the background. However, as we have mentioned before, these form fields will not affect our conclusions. On the other hand, $S_{\rm probe}$ represents the action of the probe in (\ref{dbi}). In terms of the gravitational description, $S_{\rm grav} \sim \left(1/ g_s^2 \right)$ since it originates from the closed string sector, whereas $S_{\rm probe} \sim \left(N_f / g_s \right)$, which originates from the open string sector. Clearly, the total path integral factorizes into the gravitational part and the probe part.

Following \cite{Itzhaki:1998dd}, or more recently \cite{Antonyan:2006vw}, it can be argued that stringy physics can be decoupled from a purely (super)gravity one in which one can use the gauge-gravity duality safely, as long as the curvature scale is sufficiently small compared to the string scale and the local string coupling is small enough. It is also argued in \cite{Itzhaki:1998dd}, that in the limit the local string coupling grows, one may be able to use an S-dual description or an M-theory uplift and subsequently use the corresponding gravitational description to explore the strongly coupled regime of the dual field theory. At any rate, there exists a range of energy, for which a classical gravity dual description is valid.

Now, as long as we have a classical gravitational description, we can certainly probe the geometry provided $N_f \ll N_c$. However this condition may be too naive. In reality, in order for a brane to be a legitimate probe of a background, one needs to compute the stress-energy tensor coming from the probe, denoted by $T_{\mu\nu}^{\rm probe}$, to the Einstein tensor of the background, denoted by $E_{\mu\nu}^{\rm back}$ and impose:
\begin{eqnarray}
T_{\mu\nu}^{\rm probe} \ll E_{\mu\nu}^{\rm back} \ . \label{probecond}
\end{eqnarray}
The non-trivial ingredient that enters the relation in (\ref{probecond}) is the radial-scale of the background geometry, which can override the $N_f \ll N_c$ limit. 

Now, it can also be checked, {\it e.g.}~if we place $N_f$ number of D$(p+4)$-brane probe in the near horizon geometry of a stack of D$p$-branes\cite{Itzhaki:1998dd}, the condition in (\ref{probecond}) is satisfied, at least in a non-vanishing range of the energy scale (a detailed account will appear in \cite{Kundu:tobe}). This essentially means that we can use the purely classical gravitational description and the probe sector {\it path integral} is completely determined by the classical Dirac-Born-Infeld contribution (there may be Wess-Zumino term, but it would not change our conclusions). By now, this has become a common practice in gauge-gravity duality, see {\it e.g.}~\cite{Kruczenski:2003uq, Mateos:2007vn}.

Now, let us comment on ${\cal Z}_{\rm QFT}$. Gauge-gravity duality states that the gravitational part is dual to the adjoint sector of the large $N_c$-gauge theory and the probe sector corresponds to the fundamental flavour that has been introduced in the probe limit. Likewise, we have
\begin{eqnarray}
{\cal Z}_{\rm QFT} & = & e^{i S_{\rm QFT}} \ , \\
S_{\rm QFT} & = & S_{\rm adj} + S_{\rm fund} \ .
\end{eqnarray}
Here, typically $S_{\rm adj} \sim N_c^2$ and corresponds to the adjoint sector, while $S_{\rm fund} \sim \left( N_f N_c \right)$ corresponds to the fundamental sector. At this point, one can expect that the fundamental sector is a legitimate probe in the limit $N_f \ll N_c$, however, the legitimacy of the probe limit can depend on the energy-scale of the theory. Recall, that the 't Hooft coupling is generally dimensionful, so it can set the energy-scale where back-reaction by the flavour sector becomes important. This is precisely the dual realization of the more stringent condition in (\ref{probecond}). And, as we have argued just below equation (\ref{probecond}), for a large class of examples, there exists at least a non-vanishing energy-range for which the probe limit is perfectly legitimate. For such cases, the QFT path integral also factorizes between the adjoint sector and the fundamental sector.

Now we can introduce the standard trick of thermal field theory: First Euclideanize the time direction and then compactify on it. Finally the path integral is identified with the partition function and the period of the resulting $S^1$ is identified with the inverse temperature of the system. Note, here we can impose two different periodicities for the adjoint and the fundamental sector, respectively. These two periodicities correspond to regularity condition of the Euclidean closed-string metric in (\ref{metgen}) and the regularity of the open string metric in (\ref{osmmet}). Since the modes in (\ref{fscalar})-(\ref{fspinor}) on the probe perceive the open string metric, we will focus on the latter.

Now we are ready to discuss the resulting {\it thermodynamic} description for the probe. We can generalize the proposal of \cite{Alam:2012fw} for the current situation. Let us work with the Euclidean action, which we denote by $S_{\rm DBI}^{\rm (E)}$:
\begin{eqnarray}
S_{\rm DBI}^{\rm (E)} = \N_{T_{\rm eff}} \int du \, \L^{\rm (E)} \left(a_x', u \right) \ ,
\end{eqnarray}
which can be obtained from (\ref{dbi}) after substituting the ansatz for the gauge field in (\ref{classgauge}) and then Euclideanizing the resulting action. Note that, all other constants and volume factors have been absorbed in the constant $\N_{T_{\rm eff}}$. Evidently, variation of this action yields
\begin{eqnarray}
\delta S_{\rm DBI}^{\rm (E)} = \N_{T_{\rm eff}} \left[ \left. \frac{\partial \L^{\rm (E)} }{\partial a_x'} \delta a_x \right|_{\rm horizon}^{\rm boundary} - {\rm EOM }\right] \ ,
\end{eqnarray}
where ``horizon" and ``boundary" respectively stand for the OSM horizon and the conformal boundary. Generically, we can impose $\delta a_x (\rm boundary) = 0$, and subsequently we need to subtract off the boundary contribution coming from $\delta a_x(\rm horizon)$. The boundary-terms-subtracted on-shell Euclidean action is then given by
\begin{eqnarray}
\left. S_{\rm DBI}^{\rm (E)} \right|_{\rm on-shell} = \N_{T_{\rm eff}} \left[ \int du \, \L^{\rm (E)} - j \int du \, a_x' \right] \ ,
\end{eqnarray}
where we have used the definition of $j$ from (\ref{defj}).

Thus the general proposal for the corresponding Helmholtz free energy is given by
\begin{eqnarray}
\F_H  & \equiv & T_{\rm eff} \left. S_{\rm DBI}^{\rm (E)} \right|_{\rm on-shell} \nonumber\\
          & =  & N_f T_p \int_{0}^{u_*} du \, d^p\xi \left(L_{\rm os} - j a_x' \right) \ , \label{thermosteady}
\end{eqnarray}
where $L_{\rm os}$ is given in (\ref{lagon}). We have evidently written down the expression in the full $10$-dimensional geometry. It can be noted that the conjectured free energy is essentially a Legendre transformation of the on-shell Euclidean probe action. The associated effective entropy is now given by
\begin{eqnarray}
s_{\rm eff} = - \left( \partial \F_H/ \partial T_{\rm eff}\right) \ . \label{seff}
\end{eqnarray}
Note that, the boundary term in (\ref{thermosteady}) contributes non-trivially in the entropy of the system.

Before leaving this section, let us ponder over an interesting observation. Assuming that $g_{tt}$ has a simple zero, where the closed string event horizon is located, it can be verified that 
\begin{eqnarray}
j \int_0^{u_H} a_x' \sim \left(j \cdot E\right) \tau  \ , \quad \tau \sim \log \epsilon \ ,
\end{eqnarray}
where $u_H$ denotes the closed string event horizon, $\tau$ is a typical time-scale and $\epsilon \to 0$. This time-scale --- which has a logarithmic divergence --- can be identified with the time light rays take to travel from the conformal boundary to the background event horizon. Intriguingly, this time-scale appears in an Ohmic dissipation term when the boundary term in (\ref{thermosteady}) is evaluated at the closed string horizon. By evaluating this boundary term at the OSM horizon, we are IR-safe in (\ref{thermosteady}).

On the other hand, there is a natural candidate for a Bekenstein-Hawking type area formula involving the open string data $\{\G_s, \S\}$. Recall that a probe action scales as $g_s^{-1}$, where $g_s$ is the closed string coupling. Thus, the natural area-law candidate is given by
\begin{eqnarray}
S_{\rm osm} = \left. \alpha \frac{1}{\G_s} {\rm Area} \left(\S \right) \right|_{u=u_*} \ , \label{ose}
\end{eqnarray}
where $\alpha$ is an undetermined constant and ${\rm Area}(\S)$ denotes the area of the open string metric event horizon. Also note that both $\G_s$ and ${\rm Area}(\S)$ are functions of the radial coordinate $u$. For a conformal theory beyond $d=2$, the Bekenstein-Hawking area formula indeed is proportional to the thermodynamic entropy obtained from (\ref{thermosteady}). This, however, is a simple consequence of conformal invariance. In general, (\ref{ose}) does not yield the thermodynamic entropy obtained from (\ref{thermosteady}). In fact it has a different scaling behaviour with respect to the temperature. It will be very interesting to explore and understand the physical significance of (\ref{ose}) in these general situations.

\subsection{Some Special Results}

Let us now discuss some special cases. (i) For a system with $\Phi=0$, where (\ref{seff}) and (\ref{ose}) are proportional to each other, the effective entropy is observed to obey a simple relation with the conductivity
\begin{eqnarray}
s_{\rm eff} \propto \sigma^{(m+1)/(m-1)} \ .
\end{eqnarray}
The above relation holds for $m>1$.  

(ii) If we assume that the dual field theory is Lorentz invariant, {\it i.e.} 
\begin{eqnarray}
\lim_{u \to 0} g_{tt} = \lim_{u \to 0} g_{xx} = \lim_{u \to 0} g_{ii} = \frac{1}{u^2} \ ,
\end{eqnarray}
then in the limit of strong electric field ({\it i.e.}~$E \gg TR$, where $T$ is the background closed-string metric temperature), we will get $E = 1/ u_*^2$ and subsequently the effective temperature takes a very generic form
\begin{eqnarray}
 T_{\rm eff} = \frac{E}{\pi} \left[ \frac{1+ \left.\frac{\partial \log\sigma}{\partial \log E} \right |_T}{  g_{uu}(1/\sqrt{E}) } \right]^{1/2} \ .  \label{gen2} 
\end{eqnarray}
In obtaining the above relation in (\ref{gen2}), we have used the conductivity obtained from (\ref{ustar}) and (\ref{current}). For a given physical system obeying such relations perhaps bears the possibility of a holographic dual description.

Let us now briefly discuss some more specific class of backgrounds. An AdS$_{d+1}$-Schwarzschild background is given by
\begin{eqnarray}
g_{tt} & = & - \frac{1}{u^2} \left(1- \frac{u^d}{u_H^d}\right) = - \frac{R^2}{u^4g_{uu}} \ , \\
g_{xx} & = & g_{ii} = \frac{1}{u^2} \ ,
\end{eqnarray}
where $R$ is the radius of the AdS-space. The dilaton vanishes for this background. The closed-string-background temperature is obtained to be: 
\begin{eqnarray}
T = \frac{d}{(4\pi u_H R)} \ .
\end{eqnarray}
In the large and small electric field limits, which can be obtained by setting $TR\ll E$ and $TR \gg E$, we get:
\begin{eqnarray}
T_{\rm eff} & \simeq & E^{1/2} \frac{\sqrt{2(m+1)}}{(2\pi R)} \quad {\rm when} \quad TR\ll E \ , \\
T_{\rm eff} & \simeq & T \quad {\rm when} \quad TR \gg E \ .
\end{eqnarray}
Interestingly, both expressions above are independent of $d$. The effective temperature in the intermediate regime is given by an interpolating function which can be analytically determined from (\ref{teff}). The limiting scaling behaviours are simple consequences of the underlying conformal symmetry of the system. Note that, the effective temperature obtained from (\ref{teff}) by setting $d=4$, $m=1$ and $\Phi=0$, we recover the results discussed in \cite{Sonner:2012if}.

Lifshitz geometry is another special class of geometries that corresponds to a dual field theory with non-relativistic scale invariance under $t \to \lambda^z t$, $x \to \lambda x$. Here $z$ is known as the dynamical exponent of the system. Such geometries can be viewed as infrared fixed points that describe low energy quantum criticality, which may be relevant for condensed matter systems. As $u\to 0$, the metric components take the form
\begin{eqnarray}
g_{tt} & = & - \frac{1}{u^{2z}} \ , \quad  g_{uu} = \frac{1}{u^{2}} \ , \\
g_{xx} & = & g_{ii} = \frac{1}{u^{2}} \ , \label{lifback}
\end{eqnarray}
where, for convenience, the radius of the background is set to unity. For such a geometry, in the limit $E\gg T$, the location of the OSM event horizon obeys $E = 1 / u_*^{1+z}$ and the analogue of the relation in (\ref{gen2}) takes the following form
\begin{eqnarray}
 T_{\rm eff} = \frac{E(1+z)}{2\pi} \left[ \frac{1+ \left.\frac{\partial \log\sigma}{\partial \log E} \right |_T}{  g_{uu}(E^{-1/(1+z)}) } \right]^{1/2} .  \label{gen2lif}  
\end{eqnarray}
For an exact Lifshitz geometry, where the entire bulk is defined by (\ref{lifback}), the effective temperature is given by
\begin{eqnarray}
T_{\rm eff} = \frac{\sqrt{(1+z)(m+z)}}{2 \pi} E^{z/(z+1)} \ .
\end{eqnarray}
A generalization of the Lifshitz background is the so called hyper-scaling violation geometry, for which similar relations can be obtained as well.

\subsection{Fluctuation-Dissipation Relations}

One of the profound consequences of the open string equivalence principle is the existence of a fluctuation-dissipation relation involving the effective temperature. In a system at thermal equilibrium, fluctuation-dissipation relations connect the linear relaxation response of the system to a small perturbation around the equilibrium configuration. In \cite{Sonner:2012if} such a relation was obtained for the current noise in a $(2+1)$-dimensional theory in a steady-state. Because of the open string equivalence principle, here we argue that {\it all open string degrees of freedom} in a holographic theory obey such a relation with respect to $T_{\rm eff}$. This claim can be explicitly verified by calculating the boundary Schwinger-Keldysh two-time correlator using the holographic description, as outlined in \cite{Herzog:2002pc}.

In a quantum field theory, the Schwinger-Keldysh propagator corresponds to the contour-ordered correlation function. Following the notations in \cite{Herzog:2002pc}, we get   
\begin{eqnarray}
G_{11} \left(t, \vec{x} \right) & = & - i \langle T_{\rm time} \, \O_1 \left(t, \vec{x} \right) \O_1(0) \rangle \ , \\
G_{22} \left(t, \vec{x} \right) & = & - i \langle \bar{T}_{\rm time} \, \O_2 \left(t, \vec{x} \right) \O_2(0) \rangle  \ .
\end{eqnarray}
Here $T_{\rm time}$ denotes time-ordering and $\bar{T}_{\rm time}$ denotes reverse time-ordering. The subscripts $1,2$ correspond to the Minkowski part of the contour. On the other hand, the advanced and the retarded Green's functions are also related to the Schwinger-Keldysh correlators {\it via}
\begin{eqnarray}
G_R (k) & = & \int dx e^{- i k x } G_R (x) \ , \\
G_R \left(x - y\right) & = & \theta \left(x^0 - y^0 \right) \langle  \left[ \O(x) , \O(y) \right] \rangle \ , 
\end{eqnarray}
From the above definitions it follows that
\begin{eqnarray}
G_{\rm sym} (k) & \equiv & \frac{1}{2} \left(G_{11}(k) + G_{22}(k) \right) \nonumber\\
& = & i \coth \left(\frac{\omega}{2 T}\right) {\rm Im} G_R(k) \ , \label{flucdissholo}
\end{eqnarray}
where $T$ is the corresponding temperature. Taking $\omega \to 0$ limit we obtain the fluctuation dissipation relation, which gives
\begin{eqnarray}
\lim_{\omega \to 0} G_{\rm sym} = 2 T \left( - \lim_{\omega \to 0} \frac{{\rm Im} G_R}{\omega} \right) \ ,
\end{eqnarray}
where the LHS represents the fluctuation and the bracket-ed term in the RHS represents the dissipation. In AdS/CFT the relation in (\ref{flucdissholo}) emerges from the Schwinger-Keldysh formalism, as demonstrated in \cite{Herzog:2002pc}.

The Schwinger-Keldysh computations consist of analyzing analyticity properties of various modes on the Kruskal patch of the corresponding bulk spacetime. For a typical black hole geometry, fluctuation-dissipation relations physically arise from the underlying black hole thermodynamics. For the modes in (\ref{fscalar})-(\ref{fspinor}), this underlying thermodynamics is governed by the open string metric event horizon and thus a corresponding fluctuation-dissipation result is imminent, where the temperature in (\ref{flucdissholo}) is replaced by the effective temperature $T_{\rm eff}$.

\section{Conclusions}

In this article, we have argued that for the holographic probe systems at NESS the physics remains thermal with an effective temperature. The corresponding particle distribution functions $n_\epsilon (T_{\rm eff})$ at an energy $\epsilon$ are expected to be given by a Bose-Einstein or a Fermi-Dirac statistics involving this effective temperature. The governing principles are the existence of an open string metric event horizon and an open string equivalence principle. The origin of this effective temperature is rooted in the combined effect of thermal fluctuation from the background and a Schwinger pair production which sets off the NESS. We further note that the thermal nature of a similar system was noticed earlier in \cite{Albash:2007bq, Erdmenger:2007bn} in describing phase transitions, and in \cite{Kim:2011qh}, where it was argued that the charge transport properties on the probe can be obtained from a membrane-paradigm-type description on the open string metric event horizon, further adding to the claim of the theme of this article.

In \cite{Das:2010yw} similar conclusions were arrived at using probe branes rotating along the compact directions, which inherits an event horizon on the worldvolume. It is known that for probe brane systems, T-duality symmetry of string theory transforms a spatial boost of the brane to a gauge field on the worldvolume. Thus, our NESS is qualitatively a T-dual description of \cite{Das:2010yw}. It is interesting to note that an underlying thermodynamic principle relates two seemingly unrelated systems by translating the induced metric horizon to an open string metric horizon. This perhaps also insinuates the deep connection between geometry and thermodynamics\cite{Jacobson:1995ab} that may go beyond Einstein gravity. For a recent article involving both an electric field and a rotation on the worldvolume, see \cite{Ali-Akbari:2013tca}.

On the other hand, our analysis suggests that for systems out-of-equilibrium and at least in a steady-state, an effective thermal description emerges rather universally. This is evidently true for a probe sector in a large-$N$ gauge theory with a holographic dual, irrespective of any further details of the system. Interestingly, a thermal nature for an inherently nonequilibrium process has also been observed elsewhere: earlier in \cite{PhysRevLett.95.267001}-\cite{Karch:2010kt}, in quantum critical systems, {\it e.g.}~\cite{PhysRevLett.103.206401}, or in aging glass system\cite{Cugliandolo.97}. These observations perhaps suggest a more ubiquitous presence of an effective thermodynamic description for systems out-of-equilibrium, even outside the lore of gauge-gravity duality.

By construction, our holographic analysis is applicable for a fundamental probe sector in a large-$N$ $SU(N)$ gauge theory. More recent progress on holography has made it possible to understand gravitational dual descriptions for $O(N)$-type gauge theories, which was originally proposed in \cite{Klebanov:2002ja}. It is an intriguing question whether such an effective thermodynamics persists for an analogous NESS situation there.

Our discussions have also been limited to the probe limit. Beyond this limit, the probe (fundamental) sector and the background (adjoint) undergoes heat exchange since $T_{\rm eff} > T$. Eventually, after the electric field is turned off, they will reach a thermal equilibrium. This should be realized as the open string equivalence principle merging with the closed string equivalence principle and having the same event horizon in the end. It will be very interesting to explicitly construct such a time-dependent geometric example where this physics is manifest, perhaps along the lines of \cite{Sahoo:2010sp}.

\section{Acknowledgements}

We are grateful to Sumit Das, Jacques Distler, Willy Fischler, Vadim Kaplunovsky, Hong Liu, Gautam Mandal, Bala Sathiapalan and Julian Sonner for useful discussions and encouragements about this work. This material is based upon work supported by the National Science Foundation under Grant Number PHY-0969020 and by Texas Cosmology Center, which is supported by the College of Natural Sciences and the Department of Astronomy at the University of Texas at Austin and the McDonald Observatory. AK is also supported by a Simons postdoctoral fellowship awarded by the Simons Foundation.


\begin{thebibliography}{99}
\bibitem{note2} 
There may also be a Wess-Zumino contribution; however, for our discussions this term will not change the physics. Hence we will not consider it explicitly. We also set $2 \pi \alpha' =1$.
\bibitem{comment3}
Note that this definition is up to a constant which we are not elaborating on. For our purposes, this constant is not relevant.
\end{thebibliography}
\end{document}